# Material identification using laboratory X-ray beam tracking: quantitativeness and signal-to-noise ratio requirements

**Sumera Rehman[1,a)], Ashkan Ajeer[1], Connor Darling[1], Marco Endrizzi[1], Alessandro Olivo[1], Silvia Cipiccia[1,2, a)]**

**AFFILIATIONS**

[1]Department of Medical Physics and Biomedical Engineering, University College London, Gower Street, London WC1E 6BT, United Kingdom

[2]Diamond Light Source, Harwell Science and Innovation Campus, Fermi Ave, Didcot OX11 0DE, UK

[a)] Author to whom correspondence should be addressed: sumera.rehman.21@ucl.ac.uk, s.cipiccia@ucl.ac.uk

**ABSTRACT**

Simultaneous structural and elemental characterisation of a specimen in a non-destructive manner is an instrumental approach with applications in a variety of fields including energy materials, cultural heritage and life sciences. This is routinely performed at synchrotron facilities, e.g. by combining X-ray imaging and X-ray fluorescence. In this work we describe an approach based on a monochromatic implementation of X-ray beam tracking (XBT), a multimodal imaging technique compatible with standard laboratory sources. Monochromatic XBT gives simultaneous access to quantitative absorption and phase properties of the sample, which are related to the atomic number and the electron density respectively: their combination allows for material discrimination. Here we focus on investigating the effect of the signal-to-noise ratio on the quantitativeness of the results, hence on the elemental identification. We present an XBT experiment performed using a standard X-ray laboratory source to identify the composition of three different test samples made out of Ag, Fe and Cu. These specific materials were selected as relevant to archaeological studies e.g. when specimen buried for centuries are in contact with the surrounding soil containing traces of these metals. We review the results, current limitations and provide guidance for future developments for structural and elemental characterisation in a laboratory setting.

**INTRODUCTION**

Inspecting materials while preserving their properties is one of the most valuable applications of X-rays across large scale synchrotron facilities, hospitals, industries, airports and universities. Various information about the sample can be extracted by using X-rays, e.g. its internal structure, its elemental and chemical composition. The combination of two or more modalities of inspection unlocks an even deeper level of understanding of the specimen and its formation/making process, which can be essential to the scientific investigation. To give only few examples, multimodal approaches have been proven crucial in environmental science e.g. to identify the origin of debris from the Fukushima nuclear accident [1], as well as in cultural heritage e.g. for investigating the degradation of pigments in paintings [2].

X-ray absorption imaging and X-ray fluorescence (XRF) are typically used for imaging and elemental characterisation, respectively. The former consists of recording the shadow cast by an object due to the attenuation of the X-ray beam in the object itself. The latter relies on the excitation of characteristic secondary X-ray emission, which is element-specific. This method has recently been applied to investigate the composition of ancient Greek medicines in powder form and used to discern genuine ingredients from contaminants coming from either burial conditions or from the tools used in the making process [3, 4].

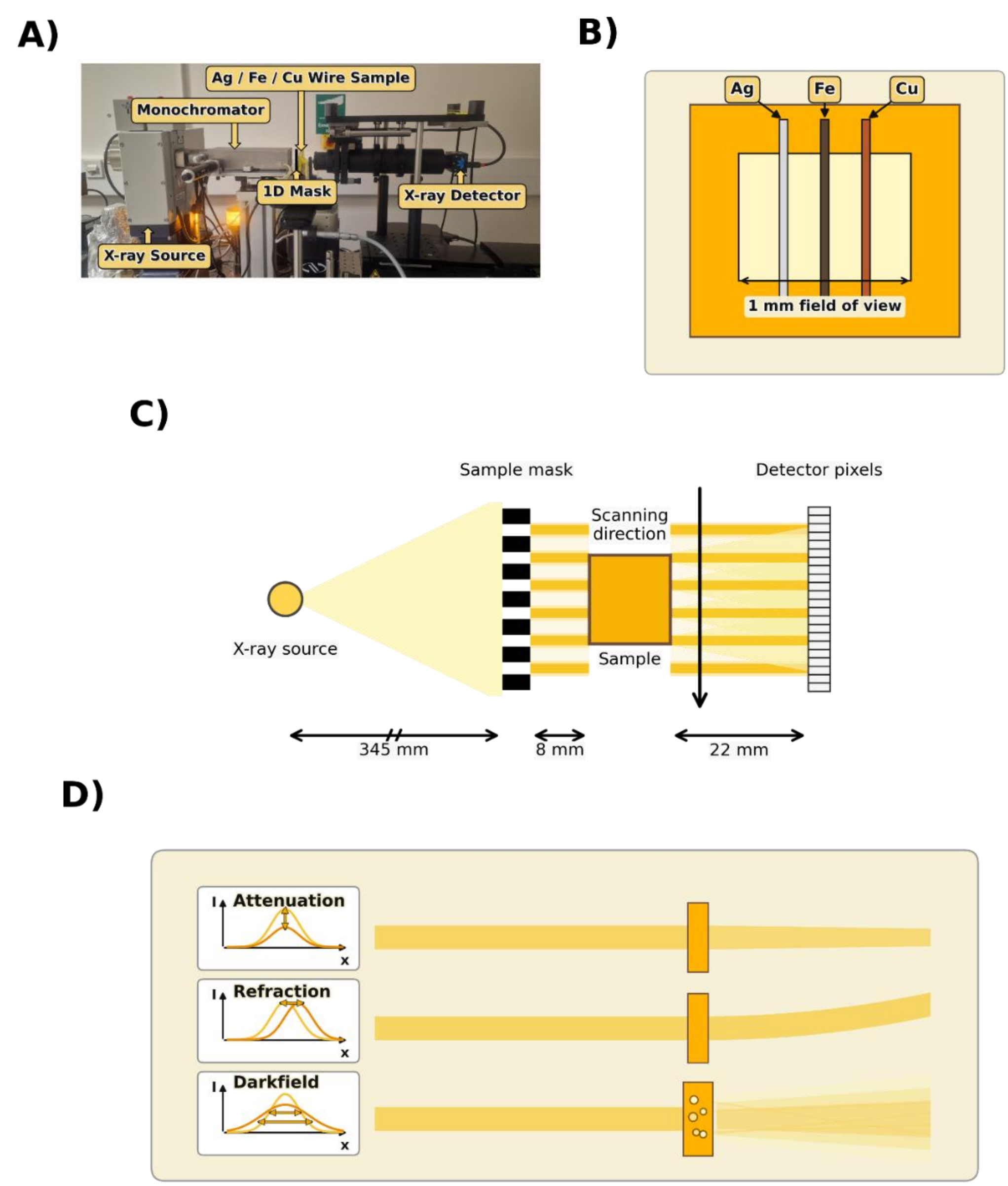

Figure 1: Overview of the experimental configuration and beam tracking principle. (A) Photograph of the NXCT X-ray beam tracking setup at UCL, showing the source, monochromator, 1D mask, Ag/Fe/Cu wire sample, and detector. The mask position is indicated by a black line. The source-mask distance was 345 mm, the mask-sample distance 8 mm, and the sample-detector distance 22 mm. Scanning was performed perpendicularly to the mask apertures. (B) Wires arranged approximately parallel to the mask apertures and positioned adjacent to one another along the scan direction. (C) Schematic of the one-dimensional beam tracking geometry. Geometry not to scale. (D) Single-beamlet interaction with the sample and the corresponding one-dimensional detected intensity profiles. Three different contrast channels can be extracted from a single image due to each type of interaction happening simultaneously.

Alternatively to XRF, it is possible to extract elemental information using only X-ray absorption imaging, by harnessing the drastic change in absorption properties of an element around its characteristic absorption edges. This can be performed by acquiring absorption images above and below the absorption edge of interest (dual energy-DE method) with a monochromatic X-ray beam [5, 6], or using a broadband radiation [7] in combination with either energy dispersive optics or energy resolving detectors. By comparing the images acquired about the absorption edge of an element, its distribution within the sample can be mapped.

The above imaging approaches use solely the absorption information. When X-rays pass through matter they are not only absorbed, but also deflected due to the interaction of the photons with the atomic electron cloud. The amount of deflection, related to the phase shift induced by the sample, depends on the spatial variations of the electron density in the sample. Qi et al. [8] proposed to perform material identification by combining absorption and phase information and they demonstrated this using a broadband laboratory X-ray source, and measuring the phase and absorption (PA) induced by the sample via grating based phase contrast imaging. Building upon [8], Buchanan and collaborators [9] showed that the PA approach is particularly valuable for low energy X-ray settings, where the broadband DE method tends to fail. They did so using the edge illumination technique [10] which, like the gratings method used by Qi et al., is a laboratory-compatible phase contrast imaging method.

A necessary step to consolidate the use of the PA method in a laboratory is to understand the effect of noise on the material identification capabilities. Indeed, standard laboratory X-ray sources offer a flux which is several orders of magnitude lower than synchrotron facilities, and the optimisation of its use is essential in any lab-based application. In this work we aim to do so and, by building on previous work by Qi *et al*. and Buchanan *et al*., we explore the robustness of the PA method for material identification against noise. We tackle the problem by using a phase contrast imaging method called X-ray beam tracking (XBT), which is an implementation of the edge illumination method where the second mask is replaced by a high resolution detector [11]. XBT consists in structuring the incoming X-ray beam with an absorbing pre-sample mask which can be 1D (periodic grid) [12] or 2D (periodic pinhole matrix) [13], to generate an array of beamlets whose positions and intensity distributions are measured with a detector having a resolution high enough to resolve them. A schematic of a typical BT setup is shown in Figure 1-C. By comparing the beamlets' intensity patterns with and without the sample, it is possible to retrieve three channels of information in a single acquisition, namely: the sample absorption, which is inferred from the decrease in intensity of the beamlets, the sample induced phase shift, which is related to the shift of the beamlets, and the dark-field signal, arising from scattering from sub-resolution features in the sample, related to the widening of the beamlets (see Figure 1-D). XBT can be performed in two modalities: static (also called single-shot) and scanning mode (also called dithering mode). In the static mode, as the name suggests, both the sample and mask are kept in the same position for the whole acquisition. As a consequence, some parts of the sample are shadowed by the absorbing parts of the mask and never 'seen' by the X-ray beam. Conversely, in the dithering mode, the sample or mask are scanned in order to illuminate the whole sample. In the static case, the resolution of the technique is determined by the mask period, while in dithering mode it can be as low as the mask aperture [14] if the scanning step is equal to half the aperture size.

To investigate the effect of noise on the PA approach, we performed XBT in dithering mode with a monochromatic X-ray beam: by varying the photon statistics in a controlled way, we assessed the effect of the signal-to-noise ratio (SNR) on the material discrimination. We collected 2D images of three different wires of known materials, Ag, Cu and Fe. These materials have been specifically selected as they are metals common in ore and soil, and therefore highly relevant in archaeological studies where the specimens may have been buried for thousands of years and contaminated by the surrounding environment [3, 4, 15].

## METHOD

*Experimental setup:*

All the experiments were performed at the UCL branch of UK's National X-Ray Computed Tomography facility (NXCT) [16]. The XBT experimental setup is shown in Figure 1-A. The X-ray source is a molybdenum rotating anode X-ray source, operated at 50 kV and 24 mA. The characteristic Mo 17.44 keV k-alpha line was selected using a multilayer flat monochromator. The X-ray beam was incident on a one-dimensional mask positioned 345 mm from the source. The mask consisted of a self-standing 35 μm thick Au grid with 2 μm apertures and 19 μm period, producing an array of narrow beamlets that propagated toward the sample. The sample was placed 8 mm downstream of the mask, and the detector 22 mm downstream of the sample. The detector was a Teledyne Moment CMOS with a pixel size of 4.5 μm, coupled with a 10x objective, giving an effective pixel size of 0.45 μm. The three metal wires were mounted on a plastic 3D printed frame within a 1 mm field of view so that they could all be imaged in a single XBT scan. The wires were aligned quasi-parallel to the mask apertures and positioned side-by-side along the scan (Figure 1-B).

Data were acquired using a dithering procedure: the sample was translated laterally to cover the magnified mask period as seen by the sample. The measured magnification of the mask at the sample was 1.02 implying that, to achieve a resolution equal to the mask aperture, a step size of 1.02 μm and a total of 19 steps would have been required to cover the full mask period. However, due to the scanning motor accuracy, the step size was effectively 1.05 μm and, therefore, the scan covered a range 3% larger than the magnified mask period at the sample. Since the relative difference is of only a few percentage points, we do not expect nor observe any effect on the retrieved data (see Results section). At every dithering position, 40 frames were recorded with an exposure of 10 s each. The average signal per frame corresponded to approximatively 120 counts per pixel above background.

*Image retrieval approach:* retrieval was performed using a moments-based approach [17]. Within each beamlet window, the zeroth spatial moment provides the integrated intensity, the first moment provides the beamlet centroid position.

The transmission was obtained from the ratio of zeroth moments, and the beamlet displacement from the difference between first moments. Beamlet displacements $\Delta x$, were converted into refraction angles using the small-angle approximation,

$$\alpha = \frac{\Delta x}{z}, \tag{1}$$

where $z$ is the sample-detector propagation distance. In XBT, it is also possible to access the dark-field signal [18, 19] via the second moments. However, this information was not used for this study, as only transmission and refraction signals are retrieved and analysed.

*Signal to Noise Ratio:* the SNR was used to quantify the statistical quality of both the raw frames and of the retrieved transmission and refraction signals. For the raw data, $SNR_{Raw}$ was defined as the ratio between the maximum signal corresponding to the mask apertures and the standard deviation of the area obscured by the mask absorption grid (see Fig. S1 in supplementary material), without sample.

$$SNR_{Raw} = \frac{T_{mas_ap}}{\sigma_{mask_abs}} \tag{2}$$

Note this is unrelated to the sample and expresses the SNR of the mask visibility [20], as one of our goals is to relate it to the SNR of the retrieved sample images. The $SNR_{Raw}$ values obtained by averaging different numbers of frames, i.e. for different photon statistics, are reported in Table *I*.

Table I: Raw-data SNR ($SNR_{Raw}$) as a function of the number of averaged frames (P).

| **P** | $\mathbf{T_{mas_ap}}$ | $\boldsymbol{\sigma_{mask_abs}}$ | $\mathbf{SNR_{Raw}}$ |
|---|---|---|---|
| 40 | 7.69 | 1.47 | 5.23 |
| 35 | 7.71 | 1.55 | 4.97 |
| 20 | 7.70 | 1.59 | 4.84 |
| 10 | 7.71 | 1.65 | 4.67 |
| 5 | 7.70 | 1.78 | 4.33 |
| 1 | 7.68 | 2.18 | 3.52 |

The next step is to calculate the SNR of retrieved transmission and phase images, which will then enable us to study their relation with $SNR_{Raw}$. For the transmission signal, $SNR_T$ was defined as the ratio between the minimum transmission value within the wire and the standard deviation of the background (see Fig. S2 in the supplementary material):

$$SNR_T = \frac{T_{min}}{\sigma_{air,T}} \tag{3}$$

where $T_{min}$ is the minimum transmission measured across the wire profile, and $\sigma_{air}$ is the standard deviation in a flat region outside the sample in the transmission image.

For the refraction signal, the SNR was defined as the ratio between the peak-to-background refraction angle and the standard deviation of the background in the refraction image (see Fig. S3 in the supplementary material):

$$SNR_R = \frac{\alpha_{max}}{\sigma_{air,R}} \tag{4}$$

where $\alpha_{max}$ is the maximum refraction angle measured across the wire profile.

**RESULTS**

*Wire thickness:*

The metal wires used for this investigation were of high elemental purity (>99.9%) and had a nominal diameter of 25 μm +/- 10%. As this investigation was in 2D, to measure the actual thickness we used the transmission profile for the highest-statistics dataset (40 projections, 40P). For each wire, the 40P transmission profile was fitted using a cylindrical chord model (Figure 2):

$$t(x) = 2\sqrt{R^2 - (x - x_0)^2}, \tag{5}$$

where $R$ is the wire radius and $x_0$ its centre. The fitted diameters, i.e. corresponding central thicknesses, are listed in Table *II*. Using these thickness values, the theoretical transmission was calculated from tabulated X-ray attenuation coefficients from the Henke database [21]. The values are reported in Table II alongside the measured transmission from 40P.

Table II: Wire diameter estimation based on the cylindrical fit approach and corresponding calculated theoretical transmission compared with the retrieved transmission values for the 40P dataset.

| **Element** | **Diameter(μm)** | **Theoretical Transmission** | **Measured Transmission** |
|---|---|---|---|
| Ag | 22.5±0.2 | 0.559±0.003 | 0.564±0.006 |
| Fe | 24.8±0.2 | 0.487±0.002 | 0.542±0.007 |
| Cu | 22.7±0.3 | 0.368±0.005 | 0.455±0.008 |

While the Ag measured and calculated transmissions show a good agreement within the experimental error, giving confidence on the estimation of the thickness, this is not the case for the Fe and Cu wires. The mismatch could be due to variation in the density of the wire with the respect to the value assumed in the Henke database for the theoretical calculation. To derive the effective thickness of the wires consistent with the measured transmission *T*, we made use of the Beer-Lambert relation,

$$T = e^{-\mu t}\ , \tag{6}$$

where μ is the linear absorption coefficient, and *t* is the thickness. Let $t_{img}$ denote the geometrically fitted thickness and $T_{th}$ the corresponding theoretical transmission. For a measured transmission $T_{ref}$, the corrected thickness is,

$$t_{new} = t_{img} \frac{ln\,(T_{ref})}{ln(T_{th})}. \tag{7}$$

The effective diameters computed in this way are given in Table *III* and were used in all subsequent quantitative analysis.

Table III: effective thickness derived from the measured transmission and Beer-Lambert relation.

| **Element** | **Corrected Thickness (µm)** |
|---|---|
| Fe | $21.2 \pm 0.4$ |
| Cu | $17.9 \pm 0.5$ |

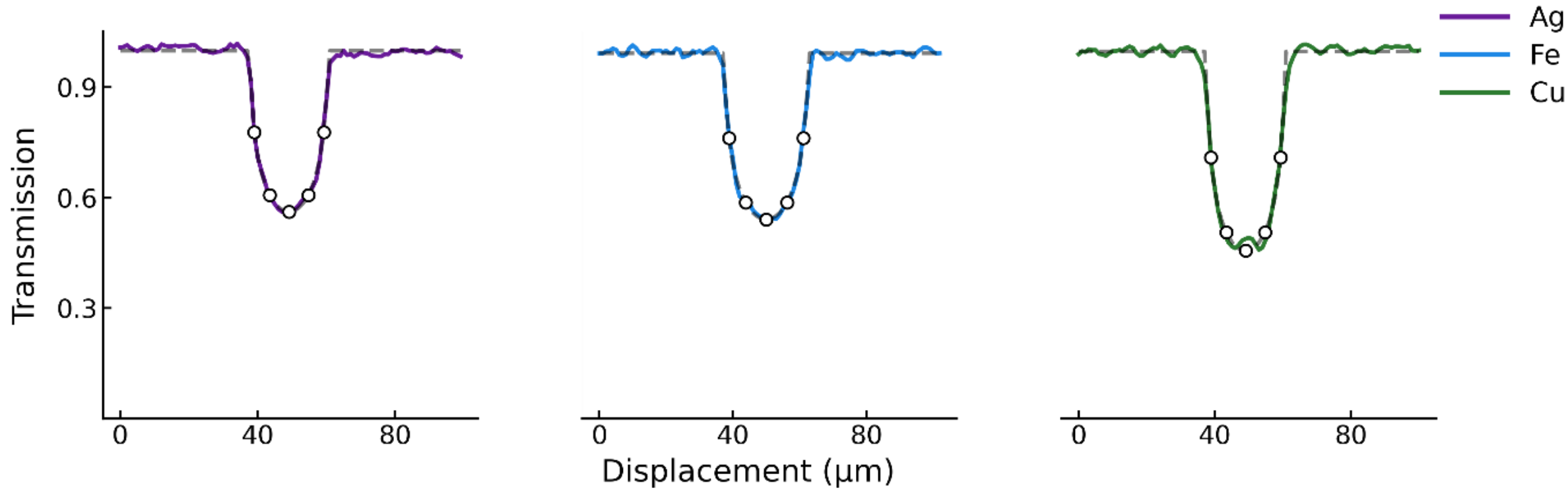

Figure 2: Cylindrical chord model fits to the 40P transmission profiles. The coloured solid lines show the measured transmission profiles and the markers indicate the sample points used in the thickness estimation. The fits were used to provide an initial estimation of the central thickness (i.e. diameter) for each wire, which was then corrected using their x-ray absorption.

To double check the correctness of the estimated effective wires thickness, we made used of the retrieved phase shift. The phase shift $\phi$ through a material of thickness $t$ is given by:

$$\phi = \frac{2\pi t\delta}{\lambda}. \tag{8}$$

$\delta$ is the refractive index decrement, given by

$$\delta = \frac{\rho_e \lambda^2 r_e}{2\pi}, \tag{9}$$

where $r_e$is the electron radius, $\rho_e$is the electron density and $\lambda$ the wavelength. The phase $\phi$ was obtained starting from the first spatial moment (centroid shift) of each beamlet and converted into refraction angle (µrad). The phase profile was then obtained by integrating the refraction signal along the scan direction. Both the refraction and the corresponding integrated phase profiles for the 40P case are shown in Figure 3. The maximum phase shift introduced by each wire was obtained from the maximum phase shift relative to the surrounding air:

$$\Delta\phi = \phi_{air} - \phi_{dip}. \tag{10}$$

The thickness values $t$ from Table III were used to compute the theoretical values for $\phi$. The resulting theoretical phase shifts based on (10) are summarised in Table IV. The measured and theoretical values are matching within the experimental error, validating the estimation of the effective wires thickness based on Beer-Lambert law.

Table IV: Theoretical phase shift calculated based on estimated thickness.

| **Element** | **Effective Thickness (μm)** | **$\rho_e$(e$^-$ /μm$^3$)** | **Theoretical $\delta$** | **Theoretical $\phi$(μrad)** | **Measured $\phi$(40P, μrad)** |
|---|---|---|---|---|---|
| Ag | 22.5 | 2.76e12 | 6.3e-6 | 12.3 | $13 \pm 1$ |
| Fe | 21.2 | 2.21e12 | 5.0e-6 | 9.4 | $10 \pm 1$ |
| Cu | 17.9 | 2.46e12 | 5.6e-6 | 8.9 | $9 \pm 3$ |

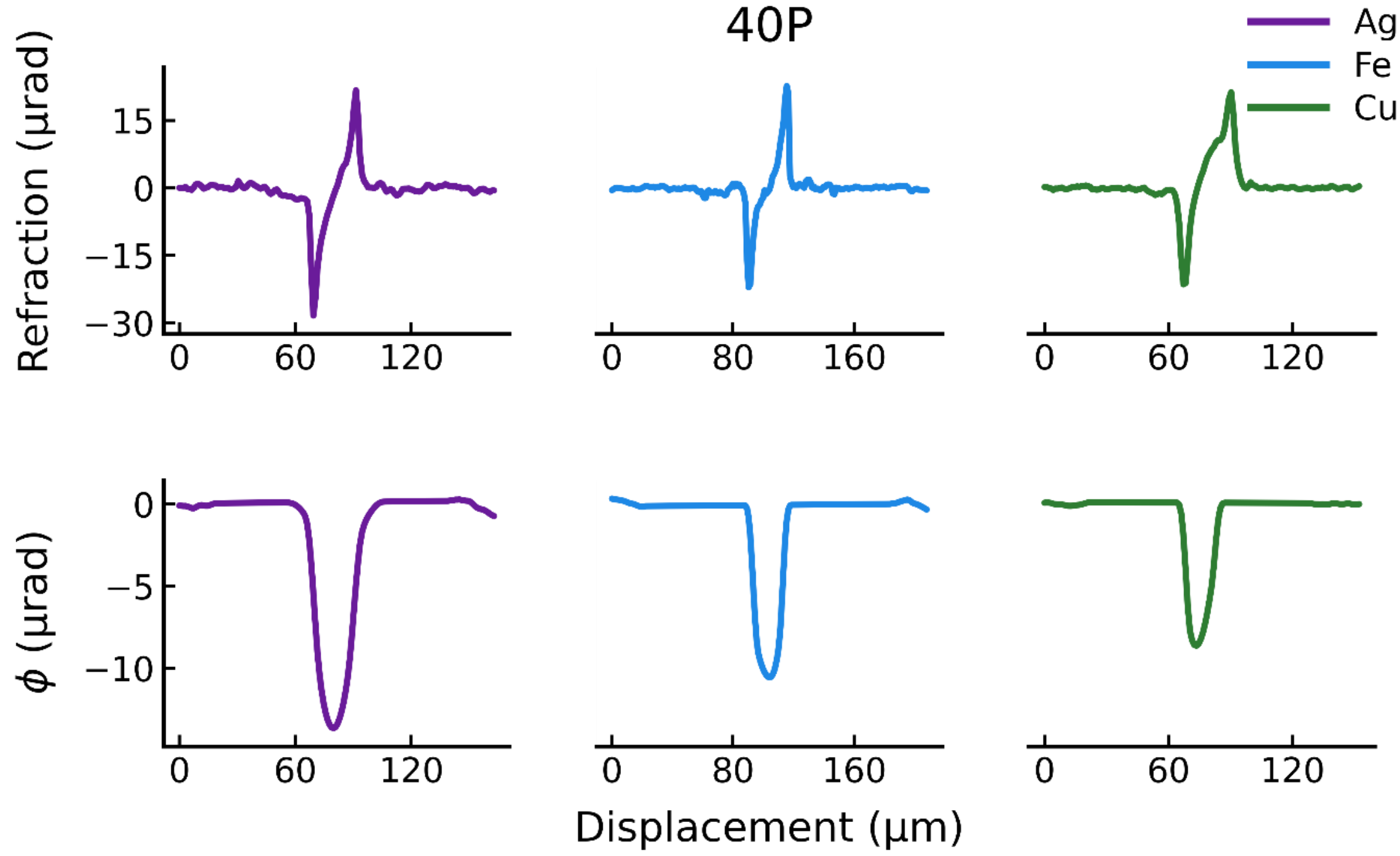

Figure 3: Refraction and phase profiles retrieved from the highest-statistics (40P) datasets for Ag, Fe, and Cu. The top row shows the refraction signal obtained from beamlet displacement, while the bottom row shows the corresponding phase profiles obtained by numerical integration of the refraction signal. The phase shift for each wire is measured as the depth of the phase minimum relative to the surrounding air.

*Effect of statistics on the quantitativeness of the measurements*

The transmission profiles and the refraction angles were retrieved for 35, 20, 10, 5 and 1 projections per dithering step, applying the same procedure as for the 40P dataset. The results are shown in Figure 4 and Figure 5. The measured values for the minimum transmission are summarised in Table *V*, while Table VI summarises the measured maximum phase shifts. For the lowest-statistics dataset (1P per dithering position), the refraction signal was extremely noisy, the phase could not be retrieved and therefore this dataset is excluded from further analysis.

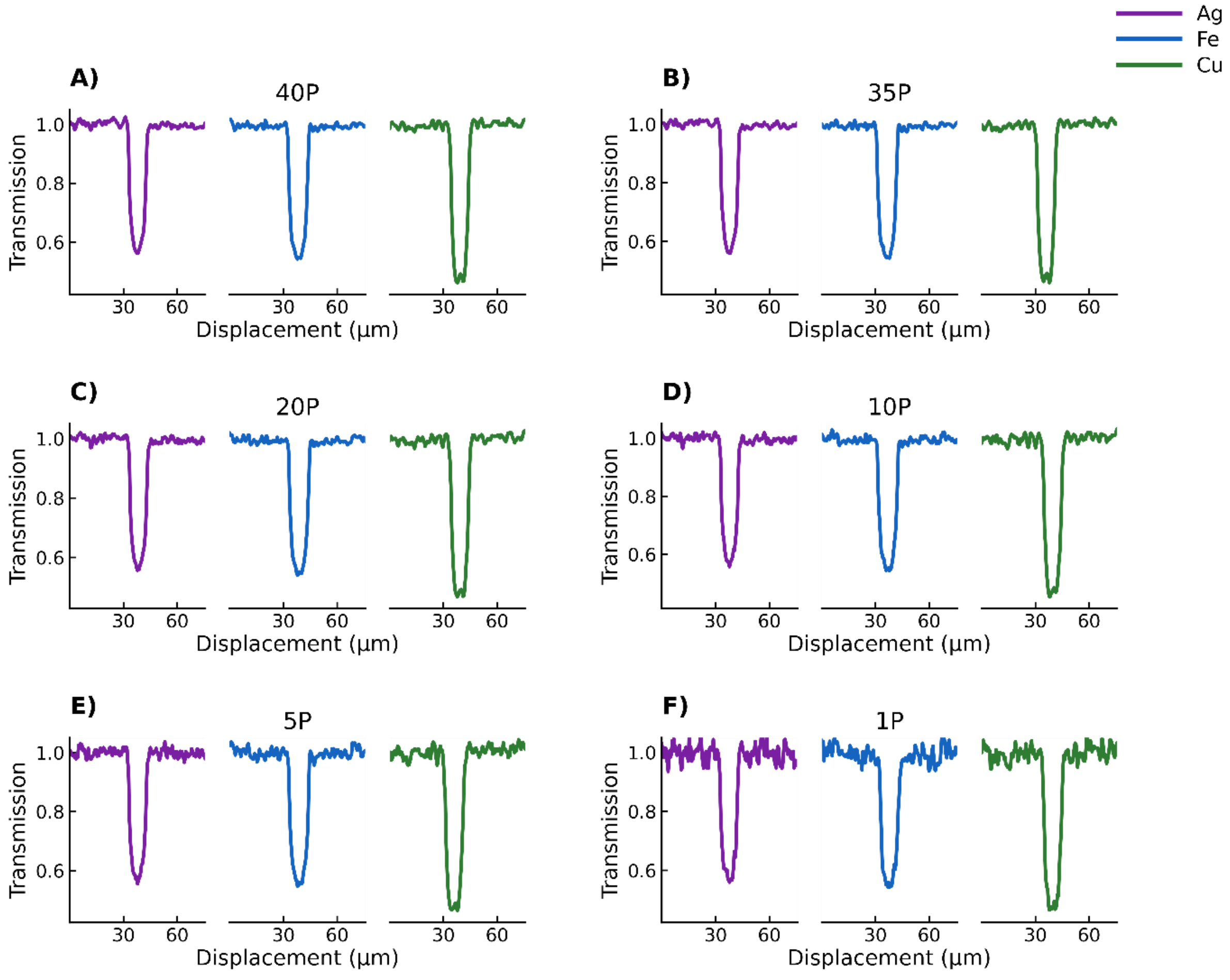

Figure 4: Retrieved transmission line profiles at different numbers of projections per dithering position: (A) 40, (B) 35, (C) 20, (D) 10, (E) 5, and (F) 1 projection. Each row shows the transmission profile across the three wires under identical acquisition conditions, illustrating the effect of reducing the number of projections i.e. the x-ray statistics.

Table V: Measured minimum transmission and SNR for the three wires for different the number of projections (P).

| P | Ag | SNR | Fe | SNR | Cu | SNR |
|---|---|---|---|---|---|---|
| 40 | 0.564±0.006 | 43 | 0.542±0.007 | 44 | 0.455±0.006 | 52 |
| 35 | 0.562±0.006 | 42 | 0.541±0.007 | 43 | 0.456±0.008 | 50 |
| 20 | 0.561±0.009 | 40 | 0.542±0.010 | 41 | 0.460±0.013 | 48 |
| 10 | 0.566±0.010 | 36 | 0.545±0.012 | 36 | 0.455±0.015 | 44 |
| 5 | 0.566±0.012 | 30 | 0.550±0.014 | 30 | 0.458±0.017 | 36 |
| 1 | 0.559±0.016 | 17 | 0.542±0.014 | 17 | 0.466±0.037 | 21 |

Table VI: Experimental maximum refraction angle, SNR and maximum phase shift as a function of the number of projections (P).

| | Ag | Fe | Cu |
|---|---|---|---|

| P | Refraction (µrad) | SNR | Phase (µrad) | Refraction (µrad) | SNR | Phase (µrad) | Refraction (µrad) | SNR | Phase (µrad) |
|---|---|---|---|---|---|---|---|---|---|
| 40 | 21.52±0.37 | 53 | 13.1±1.1 | 22.54±0.29 | 56 | 10.1±1.3 | 21.14±0.22 | 52 | 8.7±3.2 |
| 35 | 21.33±0.41 | 51 | 13.0 ±1.1 | 22.23±0.32 | 55 | 10.1 ±1.3 | 21.02±0.23 | 51 | 8.6±3.3 |
| 20 | 22.33±0.46 | 46 | 12.7 ±1.3 | 21.82±0.38 | 52 | 9.7 ±1.3 | 19.22±0.36 | 47 | 8.8±3.3 |
| 10 | 22.03±0.51 | 42 | 13.1 ±1.3 | 21.92±0.43 | 51 | 11.0±2.0 | 23.91±0.42 | 40 | 9.5 ±4.1 |
| 5 | 22.44±0.53 | 39 | 12.0±1.4 | 22.11±0.59 | 41 | 11.3 ±3.1 | 23.53±0.44 | 36 | 11.5±4.2 |

To visualise the effect of the SNR on the material discrimination capability of the PA method, we plot the refractive index decrement $\delta$ of each material against the linear attenuation coefficient $\mu$ for different number of projections (see Figure 6). The attenuation coefficient $\mu$ was calculated as follow: the transmission values sampled across each wire at five positions were extracted from the profiles in Figure 4 (these values are listed in Table S1 in the Supplementary Material). For each wire the value of $\mu$ was obtained from the linearisation of the Beer-Lambert relation (6):

$$-\ln\left(T(x)\right) = \mu t(x). \tag{11}$$

as the slope of the linear regression of $t(x)$ versus $-\ln(T(x))$ using the five sampled points for each material (fit plots shown in Fig. S4 and Fig.S5 within the supplementary material section). The resulting $\mu$ values are summarised in Table VII.

A similar approach was used to derive the refractive index decrement $\delta$. Starting from the relation between phase shift and thickness of equation (10), a linear dependence between $\phi(x)$ and $t(x)$ is expected. $\delta$ is obtained from the slope of the linear fit of $\phi(x)$ as a function of $t(x)$ (fit plots are displayed in Fig. S5 in Supplementary Material). The error bars in Figure 6 were derived directly from the measured profiles via error propagation (see "Error Propagation" section in Supplementary Material for further information).

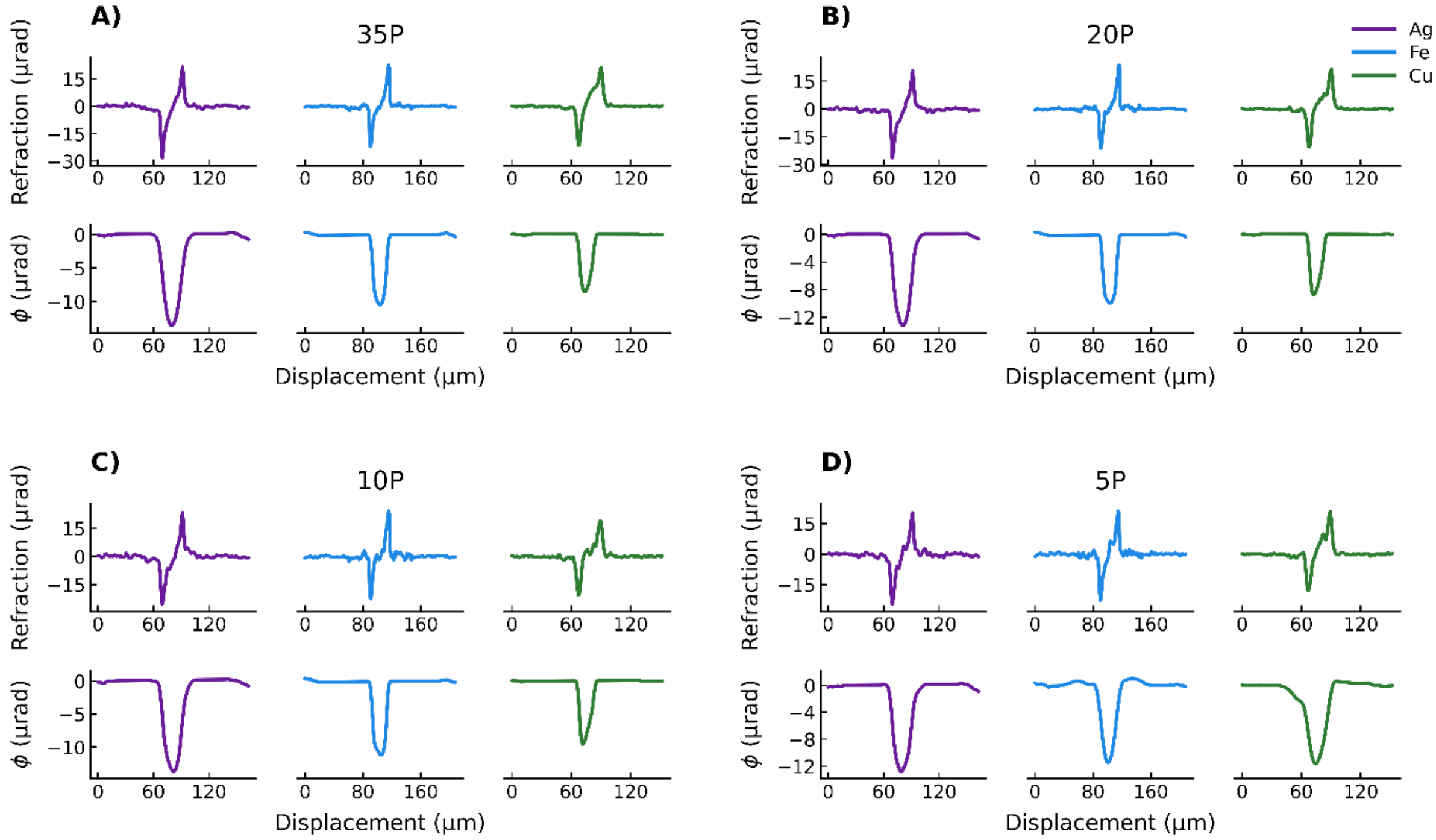

Figure 5: Retrieved refraction and corresponding integrated phase profiles at decreasing numbers of projections per dithering position: (A) 35, (B) 20, (C) 10 and (D) 5 projections. In all panels, the upper rows show the refraction signal, while the lower ones show the phase shift obtained by numerical integration.

The experimentally determined $\mu$ and $\delta$ values are summarised in Table VII. The first ("Reference") row reports the theoretical values. For comparison, the $\mu$ and $\delta$ values of nearby elements in the periodic table are given in Table VIII.

Table VII: Linear attenuation coefficient and refractive index decrement.

| | **μ_Ag (cm⁻¹)** | **μ_Fe (cm⁻¹)** | **μ_Cu (cm⁻¹)** | ***δ*_Ag** | ***δ*_Fe** | ***δ*_Cu** |
|---|---|---|---|---|---|---|
| Reference | 272.66 | 292.43 | 439.74 | 6.12e-6 | 5.07e-6 | 5.65e-6 |
| 40P | 272.66 | 292.43 | 439.74 | 6.28e-6 | 5.03e-6 | 5.60e-6 |
| 35P | 271.86 | 291.43 | 439.30 | 6.68e-6 | 5.46e-6 | 5.53e-6 |
| 20P | 271.59 | 291.40 | 433.37 | 6.48e-6 | 5.19e-6 | 5.59e-6 |
| 10P | 269.41 | 289.12 | 432.11 | 6.69e-6 | 5.87e-6 | 6.00e-6 |
| 5P | 267.47 | 287.57 | 431.83 | 6.14e-6 | 6.03e-6 | 7.29e-6 |

Table VIII: Linear attenuation coefficient of nearby elements.

| | **Pd** | **Cd** | **Mn** | **Co** | **Ni** | **Zn** | **Ga** |
|---|---|---|---|---|---|---|---|
| **μ(cm⁻¹)** | 288.87 | 235.07 | 241.32 | 361.54 | 414.64 | 382.51 | 333.02 |
| **δ** | 6.94e-6 | 4.96e-6 | 4.60e-6 | 5.64e-6 | 5.87e-6 | 4.51e-6 | 3.73e-6 |

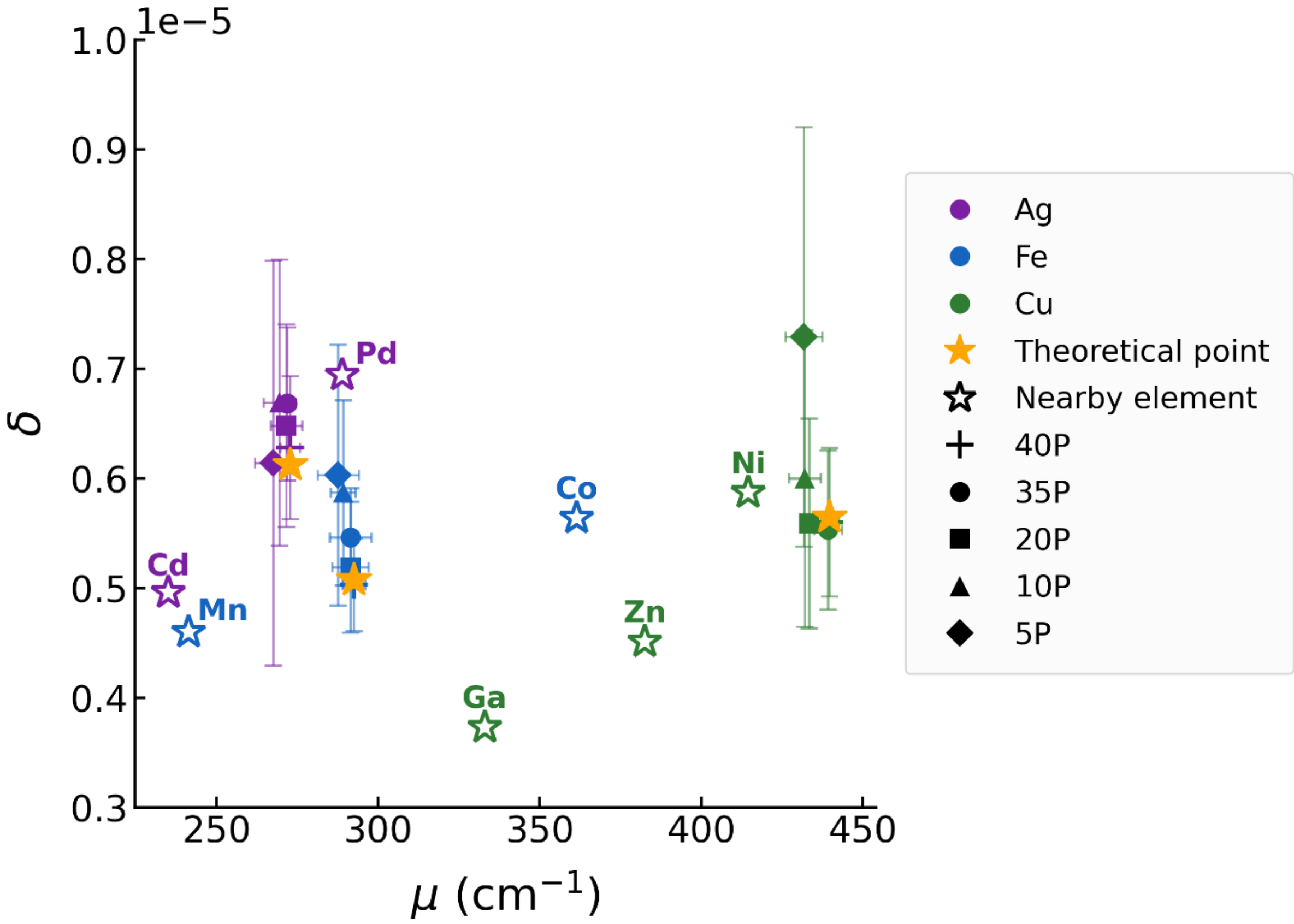

Figure 6: Linear attenuation coefficient plotted against refractive index decrement for Ag, Fe, and Cu. Coloured markers represent experimental values obtained at different projection numbers per dithering, while stars indicate theoretical reference values. Error bars are derived from transmission and phase uncertainty propagation as described in the text.

**DISCUSSION AND CONCLUSIONS**

This work investigates the effect of photon statistics on the material identification using the PA method based on XBT imaging. We have imaged three metallic wires, of materials commonly found in soil and ores and therefore relevant to the investigation of the environmental contamination specimens such as in archaeology and beyond e.g. in food science [22]. We have shown that, in high statistics conditions, both the linear attenuation coefficient $\mu$ and the refractive index decrement $\delta$ can be estimated with good agreement to theoretical values.

As expected the SNR scales with the square root of the photon numbers (see Figure S7 in Supplementary Material). Looking at the transmission values summarised in Table V, the retrieval of the transmission remains robust as the SNR decreases. As such, even at the lowest statistics investigated, the derived $\mu$ values remain consistent within propagated uncertainty with the theoretical ones. Unlike transmission, the retrieved refraction is affected more significantly by reductions in x-ray statistics. As SNR decreases, small errors in the estimated beamlet positions translate directly into larger uncertainties in the retrieved refraction angle. Converting refraction into phase requires numerical integration along the scan direction, which accumulates these fluctuations and leads to increased variability in the reconstructed phase profile. Although refraction and phase remain quantitatively reliable at moderate statistical levels, the lowest-statistics dataset (1P) cannot be retrieved. The extracted $\delta$ values therefore exhibit increasing variance at reduced SNR, consistent with the expected propagation of statistical uncertainty in phase retrieval. The higher robustness of the transmission values compared to refraction can be explained by the fact the former is derived from integrated intensity measurements (zeroth moment), which are inherently more robust to statistical noise than differential phase measurements. In Figure 7 we show how the SNR of the raw images relates to the SNR of the retrieved transmission and refraction angle, to provide guidelines for the experimental requirements to achieve elemental discrimination.

Based on the comparison of the retrieved $\mu$-$\delta$ pairs for the exemplar wire materials with the next neighbour elements (Figure 6), the PA analysis based on XBT provides a robust method for material identification. Moreover, across datasets collected at reduced statistics, the measured parameters remain clustered around theoretical values. Cu remains clearly separated in $\mu$, whereas Ag and Fe exhibit closer attenuation coefficients. In this regime, using both $\mu$ and $\delta$ reduces ambiguity between materials. Indeed looking at Figure 6, using only $\mu$ Fe and Pd could be confused, while phase enables their discrimination. From the propagated uncertainty analysis of Figure 6, the $\delta$ error bars for Ag and Fe remain separated for the 40P and 35P datasets, but begin to overlap for the 20P dataset, corresponding to a raw-data SNR of approximately 4.84 (Table I). The investigation of the effect of noise on the robustness of the methodology provides guidelines to perform reliable material identification in a laboratory settings while also investigating the structural properties of the sample.

This study is based on a two-dimensional analysis: the thickness was estimated via a cylindrical model and subsequently corrected using our pre-knowledge of the materials. In a full three-dimensional implementation, local thickness could be reconstructed tomographically, removing the need for geometric assumptions. Under such conditions, 3D $\mu$ and $\delta$ maps would enable geometry-independent material identification. Alternatively, independence from the thickness could be achieved by using the ratio $\frac{\mu t}{\delta t}$ obtainable directly from 2D projections, to discriminate the elements. Figure 8 shows this idea applied to the high statistic dataset (40P). Although the error bars are still partially overlapping with the theoretical ratio of the neighbouring elements, the plot shows the potential of an approach which would require higher statistics to be deployed effectively. However, it should also be noted that, where multiple elements overlap in a projection, the ratio will be determined by a combination of the $\frac{\mu t}{\delta t}$ ratios of the single elements, increase the complexity of this analysis and possibly rendering it ineffective.

We have focused the study on materials relevant to archaeological studies which are usually performed using synchrotron facilities to simultaneously gather structural, elemental and crystallographic information (the latter via X-ray diffraction-XRD). In this study we have shown that structural and elemental information are accessible simultaneously using conventional X-ray sources. The approach could be further extended in the future to include also the XRD analysis, e.g. by making use of the brightest laboratory X-ray sources [23], to provide a comprehensive multimodal analysis in easily accessible laboratories.

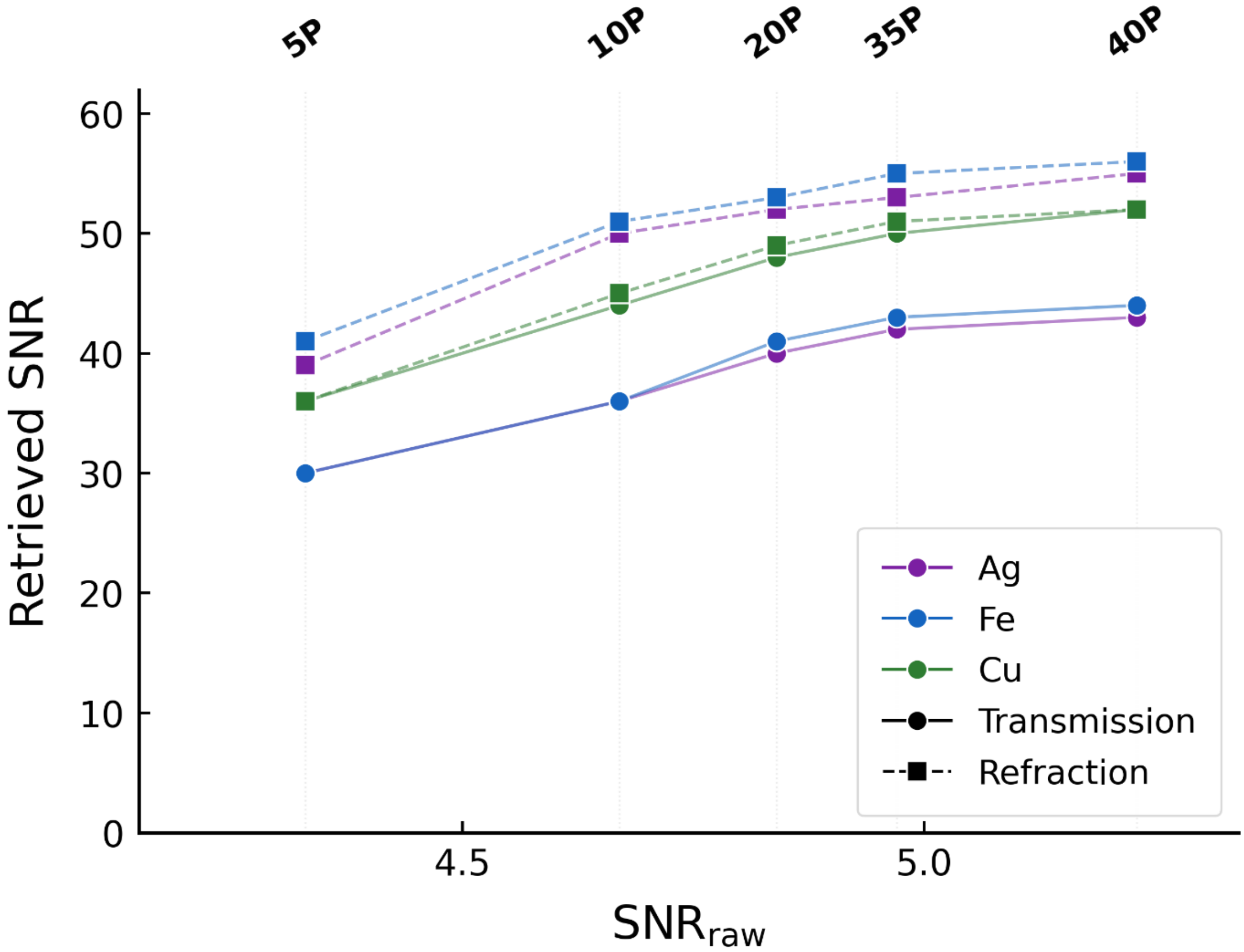

Figure 7: Experimental signal-to-noise ratio (SNR) of the retrieved transmission and refraction signals as a function of the raw-data SNR for Ag, Fe, and Cu wires at different numbers of projections per dithering position (5P-40P). Solid lines with circular markers represent transmission SNR, while dashed lines with square markers represent refraction SNR. Both transmission and refraction SNR decrease with reducing projection number; however, the refraction signal exhibits a stronger degradation, reflecting the greater sensitivity of refraction-based retrieval to reduced photon statistics. Despite this reduction, the refraction contrast remains consistently higher than the corresponding transmission contrast across all projection numbers investigated.

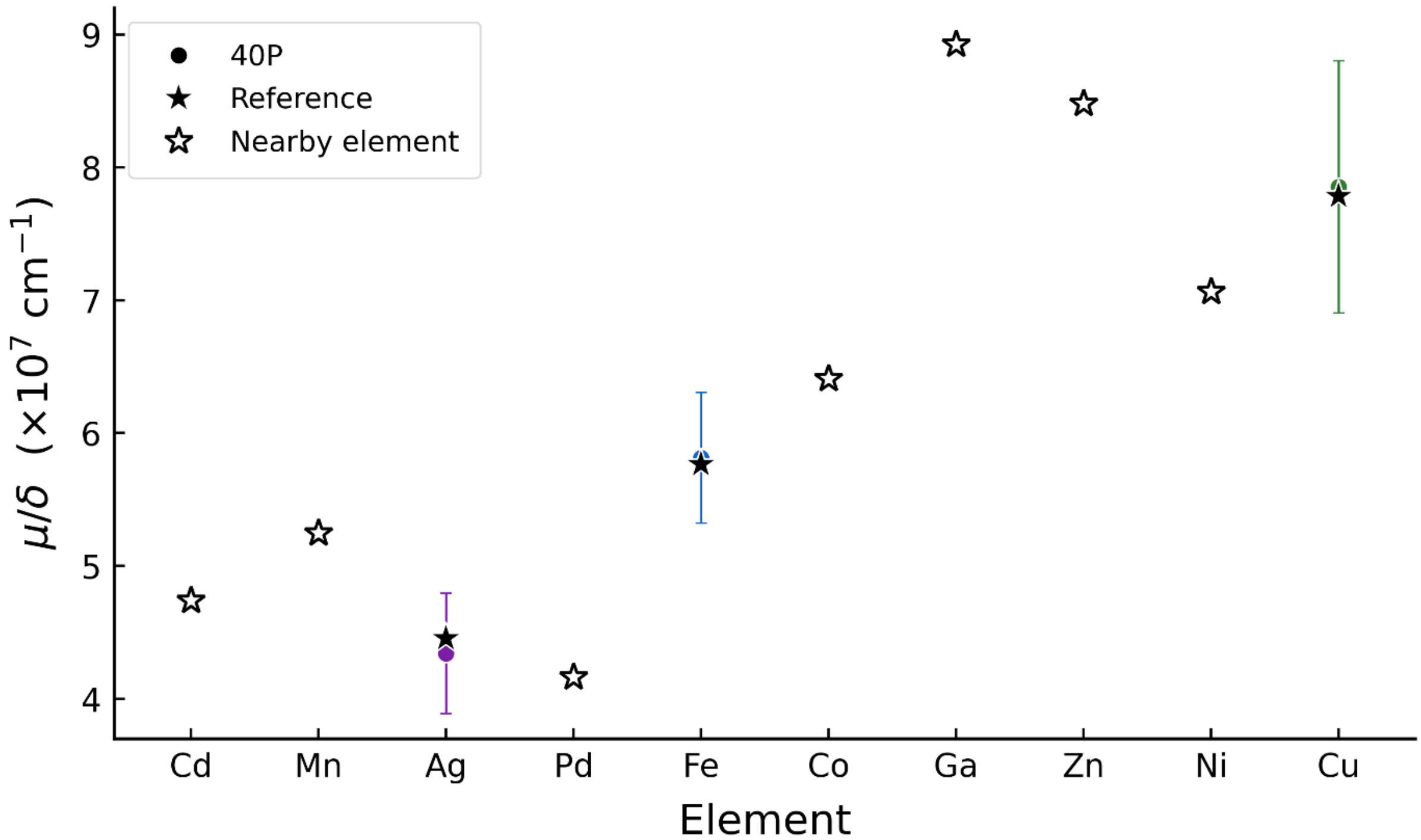

Figure 8: Ratio of the linear attenuation coefficient to the refractive index decrement (μ/δ) for Ag, Fe, and Cu and their neighbouring elements. The black stars indicate the theoretical reference values calculated from tabulated attenuation coefficients and refractive index decrements, while the coloured markers show the experimentally determined values obtained from datasets with different numbers of projections per dithering position (in this case just 40P).

## AUTHOR CONTRIBUTIONS

**S.Rehman:** Investigation (equal); Methodology (equal); Software (equal); Data curation (equal); Data interpretation(equal); Formal analysis (lead); Visualisation (lead); Writing – original draft (lead). **A.Ajeer:** Data curation (equal); Investigation (equal); Methodology (equal); Software (equal); Writing – review & editing (equal). **C.Darling:** Data curation (equal); Methodology (equal); Writing – review & editing (equal). **M.Endrizzi:** Resources (lead); Methodology (equal); Software (equal); Writing – review & editing (equal). **S.Cipiccia:** Supervision (equal); Writing – review & editing (equal), Data interpretation(equal); Funding acquisition (lead). **A.Olivo:** Conceptualisation (lead); Supervision (equal); Funding acquisition (lead); Writing – review & editing (equal).

## ACKNOWLEDGEMENTS

This work was supported by the National Research Facility for Lab X-ray Computed Tomography (NXCT). The authors are grateful to Michela Esposito for providing the Cu wire and to Carlo Peiffer for supplying the Ag wire used in the XBT experiment. We thank Harry Allan for his input with data analysis and interpretation and Alberto Astolfo for aiding with sample mount preparation.

## AUTHOR DECLARATIONS

### Conflict of interest

The authors have no conflicts to disclose.

## DATA AVAILABILITY

The data that support the findings of this study are available from the corresponding author upon reasonable request.

**SUPPLEMENTARY MATERIAL**

**SNR calculations:**

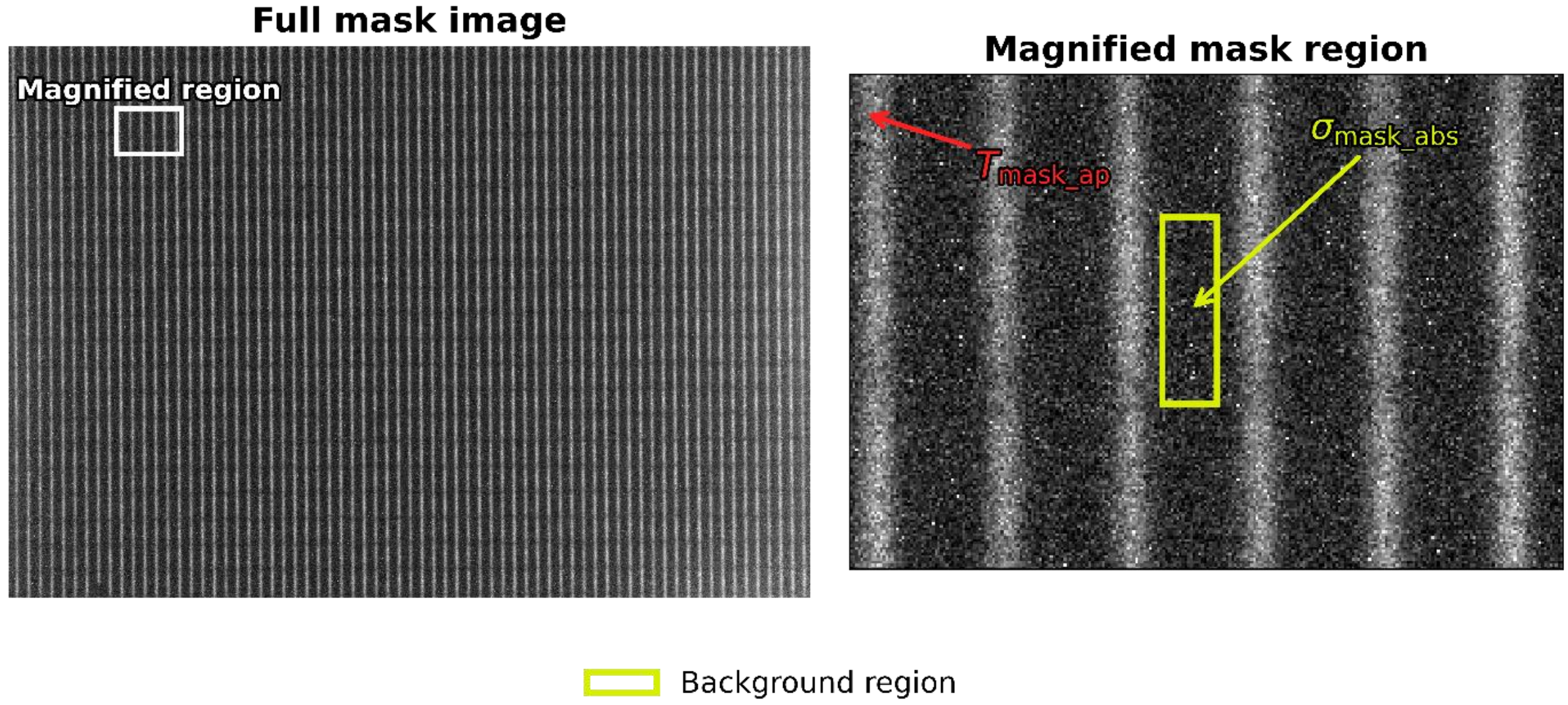

Fig. S 1: Raw mask image used for determination of the $SNR_{Raw}$. The left panel shows the full mask image, with the highlighted region corresponding to the magnified section displayed on the right. In the magnified view, the red arrow indicates the aperture transmission region ($T_{mask_ap}$), while the yellow box marks the background region used to determine the standard deviation of the mask absorption signal ($\sigma_{mask_abs}$).

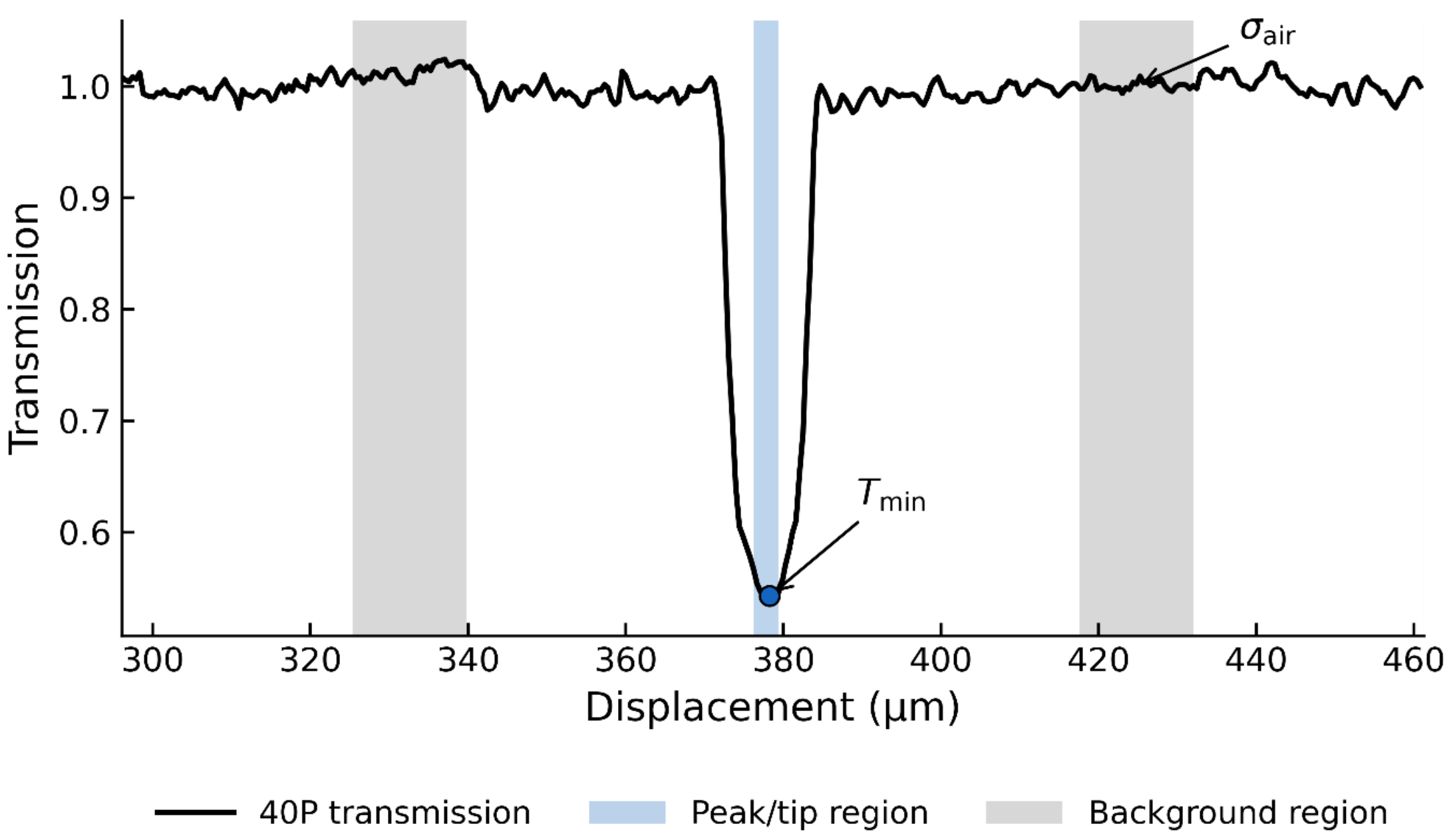

Fig. S 2: Representative transmission profile obtained from the Fe wire during the beam-tracking experiment, obtained by summing 40 frames. The shaded blue region indicates the minimum transmission value ($T_{min}$) used to define the signal amplitude, while the grey regions indicate the air/background regions used to determine the background standard deviation ($\sigma_{air,T}$). The transmission SNR was calculated as $SNR_T = T_{min}/\sigma_{air,T}$.

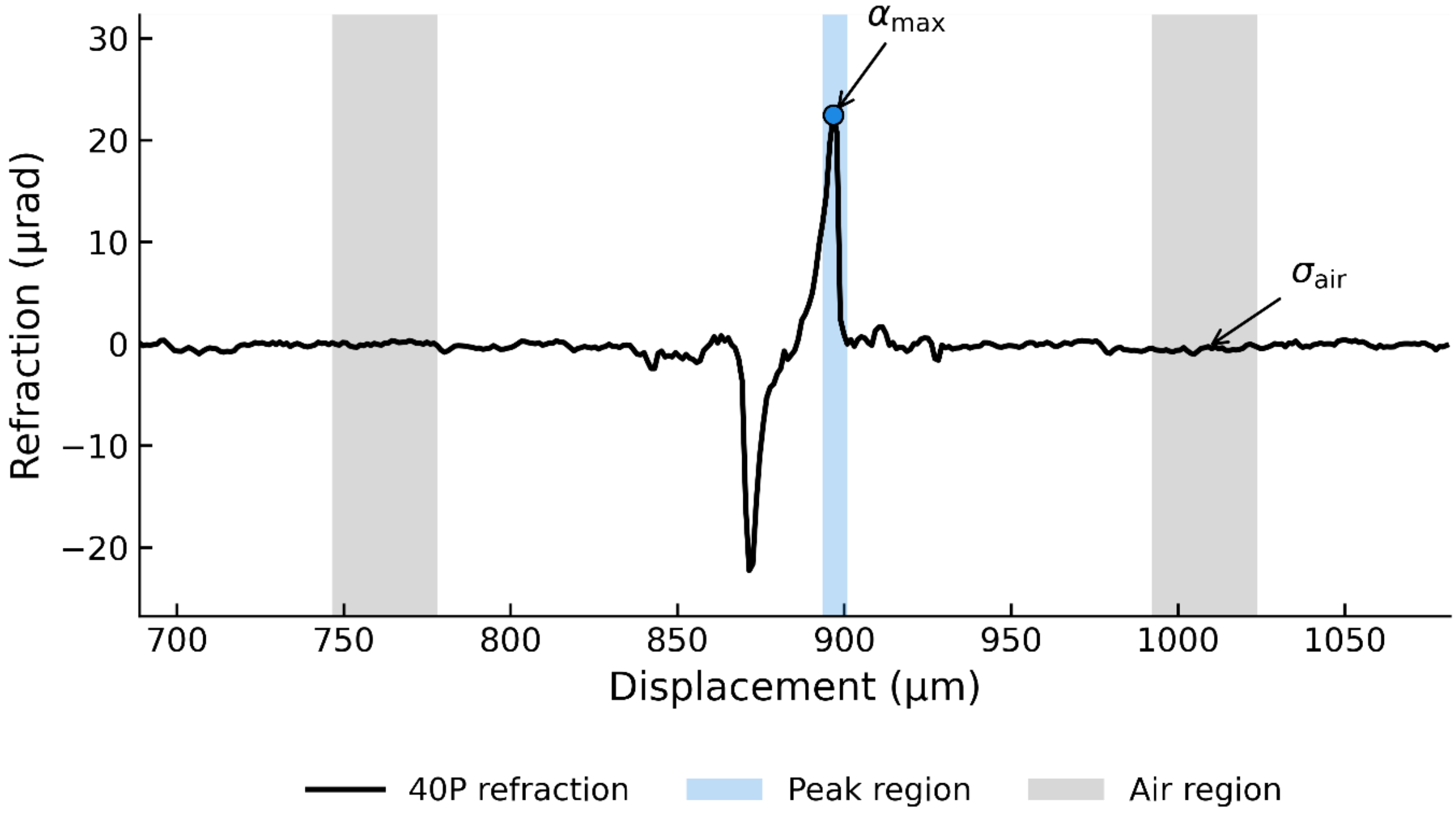

Fig. S 3: Representative 40P refraction profile obtained from the Fe wire during the beam-tracking experiment. The shaded blue region indicates the maximum refraction signal ($\alpha_{max}$), while the grey regions indicate the air/background regions used to determine the background standard deviation ($\sigma_{air,R}$). The refraction SNR was calculated as $SNR_R = \alpha_{max}/\sigma_{air,R}$.

**Determination of attenuation and phase**

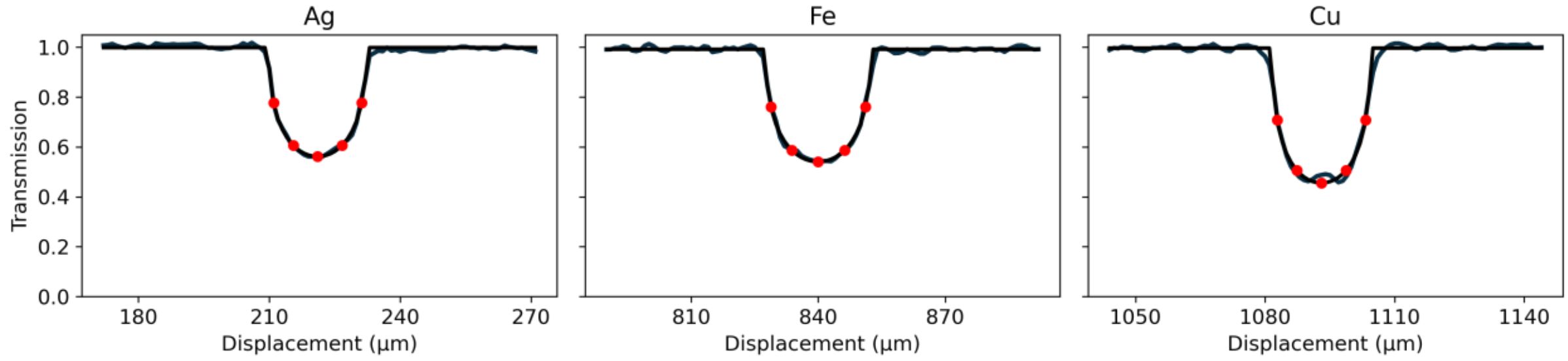

Fig. S 4: five spatial positions sampled across the wires to aid in determining the linear attenuation coefficient.

Table S 1: Transmission values sampled at the five spatial positions across each wire. These values are used to generate the data points used in the linear regressions shown in Fig. S 5.

| Element | X position | 40P | 35P | 20P | 10P | 5P | 1P |
|---|---|---|---|---|---|---|---|
| **Ag** | 210.99 | 0.7789 | 0.7770 | 0.7757 | 0.7798 | 0.7811 | 0.7778 |
| **Fe** | 828.82 | 0.7618 | 0.7614 | 0.7612 | 0.7640 | 0.7699 | 0.7574 |
| **Cu** | 1082.81 | 0.7079 | 0.7086 | 0.7110 | 0.7080 | 0.7105 | 0.7096 |
| **Ag** | 215.48 | 0.6086 | 0.6066 | 0.6055 | 0.6111 | 0.6109 | 0.6091 |
| **Fe** | 833.79 | 0.5873 | 0.5865 | 0.5873 | 0.5907 | 0.5961 | 0.5857 |
| **Cu** | 1087.37 | 0.5053 | 0.5060 | 0.5100 | 0.5054 | 0.5079 | 0.5044 |
| **Ag** | 221.09 | 0.5635 | 0.5616 | 0.5605 | 0.5664 | 0.5659 | 0.5994 |
| **Fe** | 840.00 | 0.5415 | 0.5407 | 0.5418 | 0.5452 | 0.5504 | 0.5418 |
| **Cu** | 1093.06 | 0.4550 | 0.4556 | 0.4599 | 0.4550 | 0.4575 | 0.4658 |
| **Ag** | 226.70 | 0.6086 | 0.6066 | 0.6055 | 0.6111 | 0.6109 | 0.6091 |
| **Fe** | 846.21 | 0.5873 | 0.5865 | 0.5873 | 0.5907 | 0.5961 | 0.5857 |
| **Cu** | 1098.74 | 0.5053 | 0.5060 | 0.5100 | 0.5054 | 0.5079 | 0.5044 |
| **Ag** | 231.19 | 0.7789 | 0.7770 | 0.7757 | 0.7798 | 0.7811 | 0.7778 |
| **Fe** | 851.18 | 0.7618 | 0.7614 | 0.7612 | 0.7640 | 0.7699 | 0.7574 |
| **Cu** | 1103.30 | 0.7079 | 0.7086 | 0.7110 | 0.7080 | 0.7105 | 0.7096 |

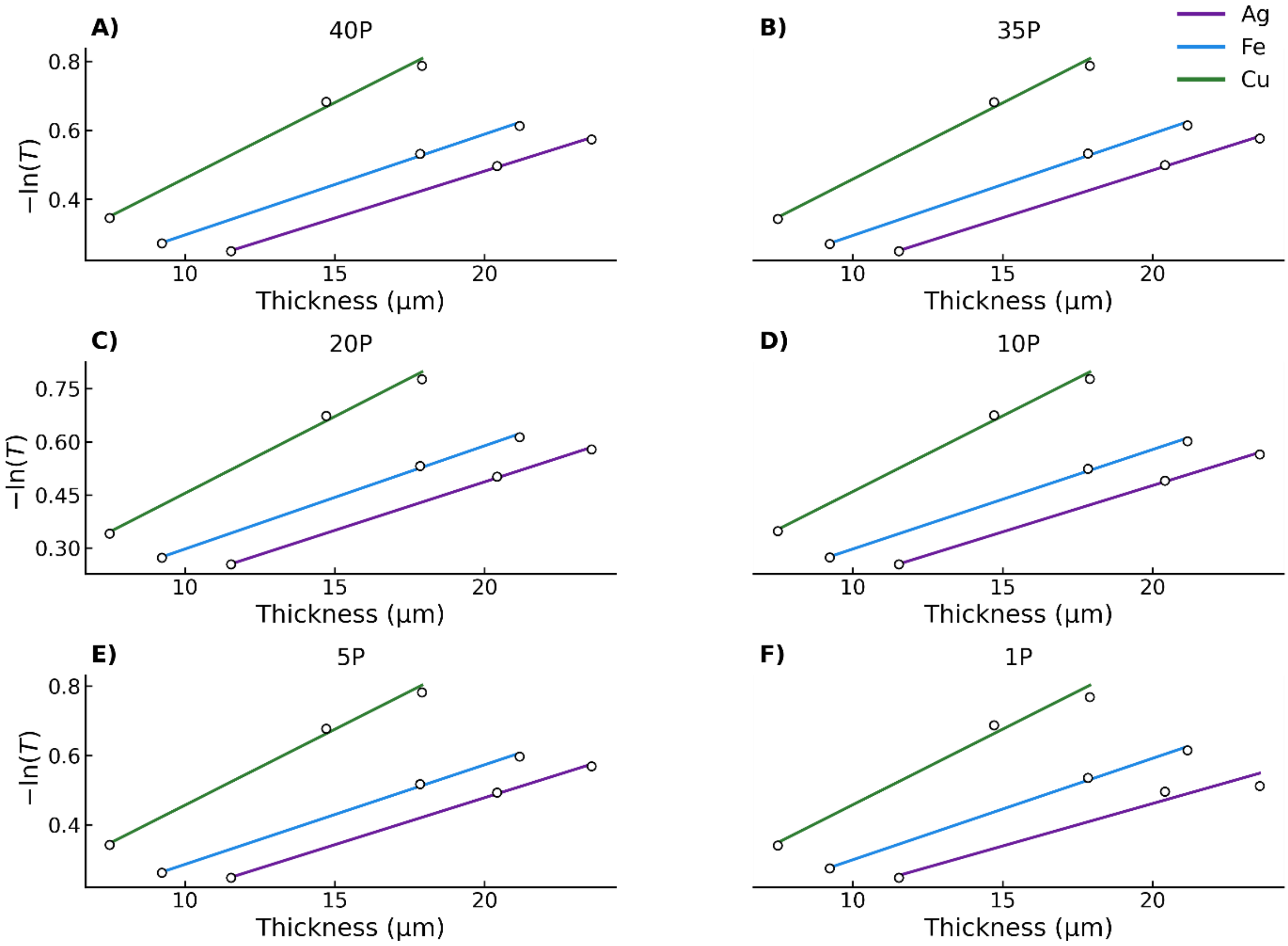

Fig. S 5: Thickness plotted against $-ln\ (T)$ for Ag, Fe, and Cu at (A) 40, (B) 35, (C) 20, (D) 10, (E) 5 and (F) 1 projections per dithering position. Solid lines indicate least-squares linear fits. The slope of each fit provides the linear attenuation coefficient μ.

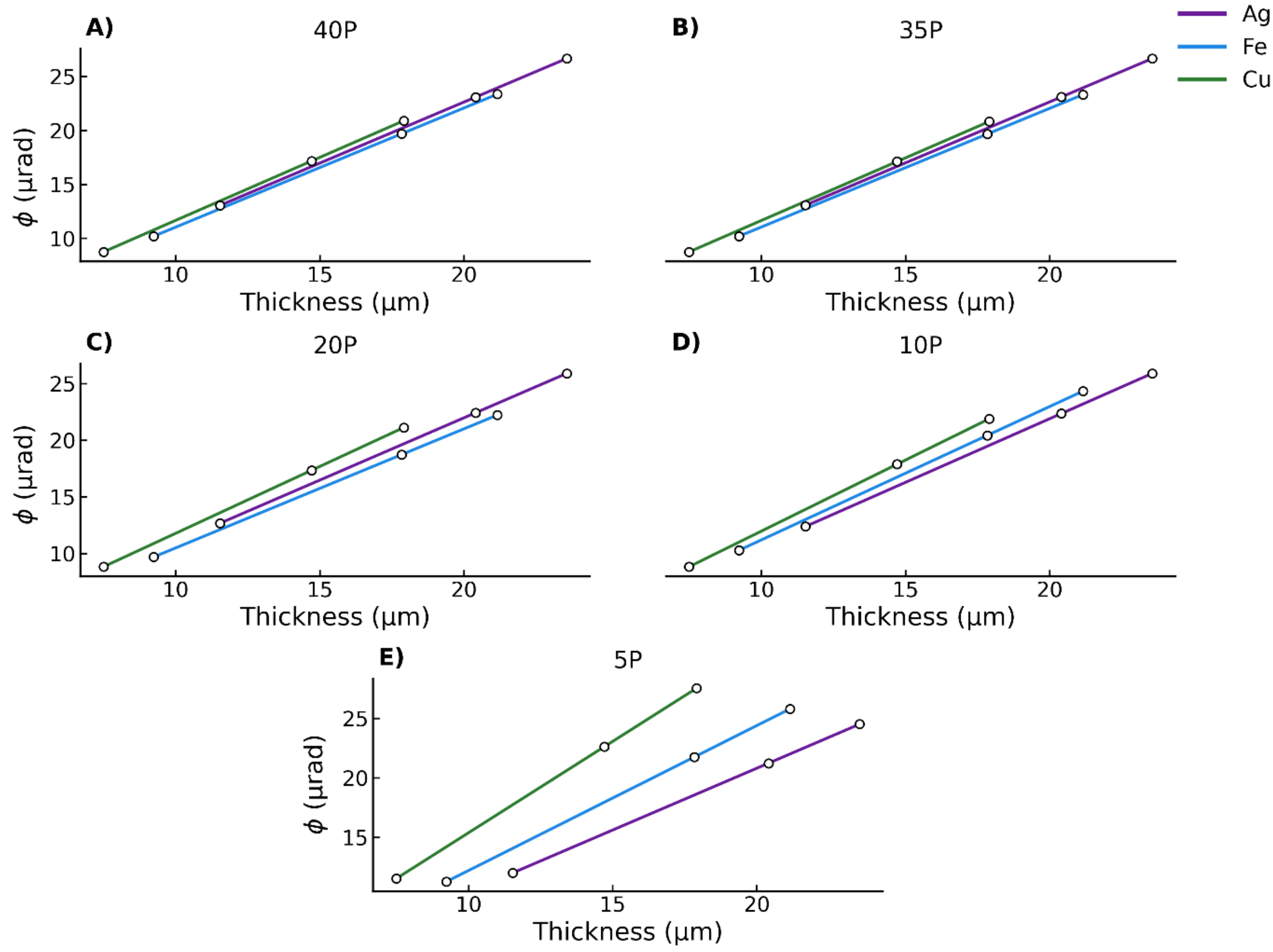

Fig. S 6: Thickness plotted against $\phi$ for Ag, Fe, and Cu at (A) 40, (B) 35, (C) 20, D) 10 and (E) 5 projections. Solid lines represent linear least-squares fits, where the slope corresponds to the phase shift ($\delta$).

**Error propagation:**

For the transmission uncertainty this is given by:

$$\sigma^2{}_A = \left(\frac{\partial A}{\partial I}\sigma_I\right)^2 + \left(\frac{\partial A}{\partial I_0}\sigma_{I_0}\right)^2,$$

which simplifies to

$$\sigma_A = \sqrt{\left(\frac{\sigma_I}{I}\right)^2 + \left(\frac{\sigma_{I_0}}{I_0}\right)^2},$$

where $A$ is the measured transmission quantity, $I$ and $I_0$ are minimum intensity within the sample and the intensity in the air, $\sigma_I$ and $\sigma_{I_0}$ correspond to standard deviations measured directly from the profiles. Phase uncertainty was calculated as:

$$\sigma_{\Delta\phi} = \sqrt{\sigma_{air}^2 + \sigma_{dip}^2}.$$

where $\sigma_{air}$ and $\sigma_{dip}$ are the standard deviations measured in the air and wire regions of the phase profile.

**Variation of SNR with photon statistics**

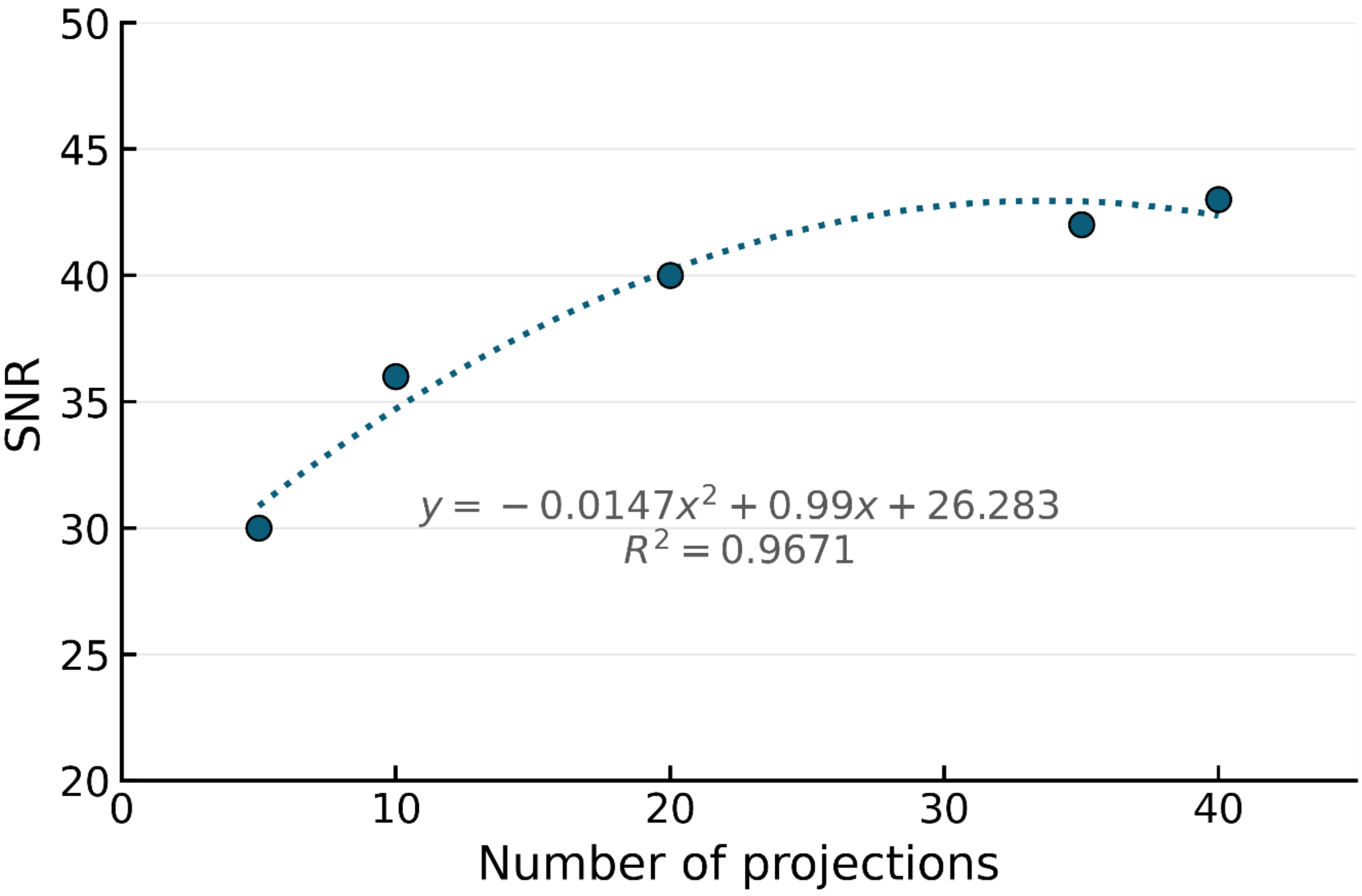

Fig. S 7: SNR Vs number of projection for the Ag wire. The higher coefficient of determination for the polynomial fitting matches the expected dependence of the SNR on the square of the photon numbers